\documentclass[aps,prd,floats,nofootinbib,preprintnumbers]{revtex4}
\usepackage{epsfig}

\begin{document}

\preprint{NUHEP-TH/04-04}

\title{Deviation of Atmospheric Mixing From Maximal and Structure in the Leptonic Flavor Sector}

\author{Andr\'e de Gouv\^ea}
\affiliation{Northwestern University, Department of Physics \& Astronomy, 2145 Sheridan Road, Evanston, IL~60208, USA}

\begin{abstract}
I attempt to quantify how far from maximal one should expect the atmospheric mixing angle to be given a neutrino mass-matrix that leads, at zeroth order, to a $\nu_3$ mass-eigenstate that is 0\% $\nu_e$, 50\% $\nu_{\mu}$, and 50\% $\nu_{\tau}$. This is done by assuming that the solar mass-squared difference is induced by an ``anarchical" first order perturbation, an approach than can naturally lead to experimentally allowed values for all oscillation parameters. In particular, both $|\cos2\theta_{\rm atm}|$ (the measure for the deviation of atmospheric mixing from maximal) and $|U_{e3}|$ are of order $\sqrt{\Delta m^2_{\rm sol}/\Delta m^2_{\rm atm}}$ in the case of a normal neutrino mass-hierarchy or of   order $\Delta m^2_{\rm sol}/\Delta m^2_{\rm atm}$ in the case of an inverted one. Hence, if any of the textures analyzed here has anything to do with reality,  next-generation neutrino experiments can see a nonzero $\cos2\theta_{\rm atm}$ in the case of a normal mass-hierarchy, while in the case of an inverted mass-hierarchy only neutrino factories should be able to see a deviation of $\sin^22\theta_{\rm atm}$ from 1.

\end{abstract}

\maketitle

\setcounter{equation}{0}
\section{Introduction}
\label{intro}
 
One of the most exciting developments in particle physics of the past six years has been the confirmation, by several different experiments \cite{SK_atm,solar, KamLAND}, that neutrinos have mass and that, similar to quarks, neutrino mass-eigenstates and weak-eigenstates mix. The existence of neutrino masses is, presently, the only solid, unambiguous evidence of physics that cannot be accommodated by the standard model of electroweak and strong interactions. However, while there is evidence for physics beyond the standard model, the amount of information available is very scarce. The two most significant facts we did learn (and would like to understand better) are: i-neutrino masses are significantly smaller than all other fermion masses ($m_{\nu}/m_e<10^{-6}$), ii-the leptonic mixing matrix is qualitatively different from the quark mixing matrix. In this paper, I'll have nothing to add regarding (i), but would like to address (ii).

The quark mixing matrix is well approximated by the unit matrix (see, for example, \cite{PDG}), {\it i.e.,}\/ the mixing angles of the CKM matrix are either small ($\sin^2\theta_{C}\simeq 0.05$) or very small ($|V_{ub}|^2\simeq 10^{-5}$). Combined with the fact that quark masses are hierarchical ($m_t\gg m_c \gg m_u$, etc), the peculiar form of the CKM matrix has led  to the belief that there is some underlying symmetry or dynamical mechanism that fundamentally distinguishes the different quark families, and naturally explains the small mixing angles and the disparate mass eigenvalues. Indeed, it seems quite implausible that the CKM matrix is representative of a ``random" unitary mixing matrix.\footnote{An attempt to quantify this statement can be found in \cite{anarchy_stat}.} 

In the leptonic sector, the situation is quite distinct. The leptonic mixing matrix is ``far" from a unit matrix, and all its entries are large, with the exception of the $U_{e3}$ element,\footnote{I define the mass eigenstates such that $m_2^2>m_1^2$, and $|\Delta m^2_{13,23}|>\Delta m^2_{12}$, where $\Delta m^2_{ij}\equiv m_{j}^2-m_{i}^2$. Weak-eigenstates $\nu_{\alpha}$ are related to the mass-eigenstates $\nu_i$ via $\nu_{\alpha}=U_{\alpha i}\nu_i$, where $\alpha=e,\mu,\tau$, $i=1,2,3$, and $U_{\alpha i}$ are the elements of the leptonic mixing matrix, also referred to as the MNS matrix, $U_{MNS}$.
Mixing angles $\theta_{ij}$ are defined as per the PDG prescription \cite{PDG}.} which is somewhat small: $|U_{e3}|^2< 0.07$ at three sigma confidence level \cite{CHOOZ,CHOOZ_bound}. It is not clear that there is, currently, strong evidence to believe that a new underlying symmetry or dynamical mechanism that distinguishes the different leptonic families is required to explain the leptonic mixing angles.\footnote{One potential criticism is that the charged lepton masses are hierarchical. While this issue will not be addressed here at all, it is easy to imagine scenarios where the charged lepton masses are hierarchical, while the leptonic mixing matrix is structureless \cite{SU(5)_anarchy}. Indeed, in some scenarios it is possible to obtain a structureless mixing matrix while charged lepton {\sl and} neutrino masses are hierarchical \cite{minimal_model}.} Indeed, it has been proposed that perhaps the neutrino-mass sector of the standard model is structureless, and that the leptonic mixing matrix is a reasonable representative of a random mixing matrix \cite{anarchy}. 

This statement has recently been verified in a more quantitative way \cite{anarchy_stat}. It was pointed out that the ``anarchical hypothesis" for the leptonic mixing matrix is consistent as long as $|U_{e3}|$ does not differ significantly from its current upper bound. Furthermore, if correct, the anarchical hypothesis provides a {\sl lower bound} for $|U_{e3}|$, namely $|U_{e3}|^2>0.011$ at the two sigma confidence level \cite{anarchy_stat}. This means that next generation searches for $|U_{e3}|^2$ using, say, neutrino ``super-beams" \cite{superbeam} or reactor antineutrinos \cite{new_reactor} can definitely test the hypothesis that the leptonic mixing matrix is consistent with a random unitary matrix.

In spite of the claims above, a significant amount of research effort has gone into determining what sort of underlying symmetries or mechanisms can explain the leptonic mixing angles and the neutrino masses \cite{models}. There are several reasons for this, and I will spell out two of them. One is the belief that, at some very high energy scale, the distinction between quarks and leptons is erased. This happens, for example, in most grand unified models. Hence, because the quark sector seems to be structured, it is likely that the same is true of the leptonic sector \cite{GUT_models} (for a counter-argument, see \cite{SU(5)_anarchy}). Another reason for believing in structure in the leptonic section is the curious fact that, currently, the atmospheric mixing angle is consistent with maximal ($\theta_{23}\simeq\pi/4$). While maximal atmospheric mixing is in agreement with the anarchical hypothesis \cite{anarchy_stat}, it can also be interpreted as evidence for specific ``textures" in the neutrino mixing matrix, and most of these textures can be related to underlying flavor symmetries. 

In this paper, I will try to address how close to maximal the atmospheric mixing angle must be if its ``maximality" is indeed related to some fundamental principle. In the next section I discuss what assumptions will be made, and define how this issue will be addressed. After that, I'll analyze different mass-matrix textures that lead to maximal atmospheric mixing at leading order, and  estimate the typical deviation of the atmospheric angle from maximal, among other observables. Discussions, a summary, and conclusions follow.

\setcounter{equation}{0}
\section{Neutrino Mass Matrices, Maximal Mixing, and Random Perturbations} 
\label{2}

Instead of concentrating on several distinct flavor models and dynamical mechanisms for explaining the pattern of neutrino masses and
 the values of the mixing angles, I'll concentrate on a bottom-up approach which imposes, at the leading order, maximal mixing in the atmospheric sector. For this purpose, I select zeroth order neutrino mass-matrices $M_0$ that, in the basis where the charged current and the charged lepton mass matrix are diagonal, have one mass-eigenstate which is 0\% $\nu_e$, 50\% $\nu_{\mu}$ and 50\% $\nu_{\tau}$. I will only consider Majorana mass matrices and do not particularly care about the ``fundamental origin" of $M_0$. The next step is to perturb $M_0$ by adding a generic ``first order perturbation" mass matrix $M_1$. See the Appendix for a definition of ``generic." As in the case of $M_0$, the origin of $M_1$ is not relevant to the considerations made here. It can be associated to the renormalization group running of the several couplings in specific models and/or heavier see-saw right-handed neutrinos and/or a different, sub-leading, source of lepton-number violation, etc.
 
 In order to make more quantitative assertions, it is important to define the size of the perturbation. This will be done by requiring that, after the perturbation is added, other features of neutrino mixing (especially the solar mass-squared difference) are properly explained. 

It proves instructive to start with two toy $2\times 2$ scenarios (for concreteness, the $\nu_{\mu}\times\nu_{\tau}$ ``subspace"). The first one is
 \begin{eqnarray}
 M_0=m_0\left(\begin{array}{cc}
 0&1\\1&0
\end{array}\right), \label{Lm-Lt} \\
M_1=m_1\left(\begin{array}{cc}
 a&c\\c&b
\end{array}\right) \label{M1_2}, 
 \end{eqnarray}
where $m_0,m_1$ ($m_0\gg m_1$) have dimensions of mass and $a,b,c$ are $O(1)$ real numbers.\footnote{It suffices to consider only real mass matrices, which is assumed henceforth. One can always deal with complex mass-matrices by diagonalizing $MM^{\dagger}=U{\rm diagonal} [|m_1|^2,|m_2|^2,\cdots]U^{\dagger}$ instead of $M=U{\rm diagonal} [m_1,m_2\cdots]U^{\top}$, and obtain, qualitatively, the same results.} It is easy to see that the eigenvalues of Eq.~(\ref{Lm-Lt}) are
$m_0,-m_0$, and its eigenvectors are $(1/\sqrt{2},1/\sqrt{2})^{\top}$ and $(1/\sqrt{2},-1/\sqrt{2})^{\top}$, respectively. While it does not concern the discussions here, it is worthwhile to mention that Eq.~(\ref{Lm-Lt}) can be interpreted as due to an $L_{\mu}-L_{\tau}$ global symmetry.

It is trivial to diagonalize Eq.~(\ref{Lm-Lt})+Eq.~(\ref{M1_2}) exactly and obtain eigenvalues and eigenvectors. It proves useful, however, to obtain approximate results via perturbation theory. Explicitly,
\begin{eqnarray}
M_0+M_1&=&m_0\left(\begin{array}{cc}1/\sqrt{2}&1/\sqrt{2}\\-1/\sqrt{2}&1/\sqrt{2}\end{array}\right)\left[
\left(\begin{array}{cc}-1&0\\0&1\end{array}\right)+\frac{m_1}{2m_0}\left(\begin{array}{cc}1&-1\\1&1\end{array}\right)\left(\begin{array}{cc}a&c\\c&b\end{array}\right)\left(\begin{array}{cc}1&1\\-1&1\end{array}\right)
\right] \left(\begin{array}{cc}1/\sqrt{2}&-1/\sqrt{2}\\1/\sqrt{2}&1/\sqrt{2}\end{array}\right), \\
&=&m_0\left(\begin{array}{cc}1/\sqrt{2}&1/\sqrt{2}\\-1/\sqrt{2}&1/\sqrt{2}\end{array}\right)\left[
\left(\begin{array}{cc}-1&0\\0&1\end{array}\right)+\frac{m_1}{2m_0}\left(\begin{array}{cc}a+b-2c&a-b\\a-b&a+b+2c\end{array}\right)
\right] \left(\begin{array}{cc}1/\sqrt{2}&-1/\sqrt{2}\\1/\sqrt{2}&1/\sqrt{2}\end{array}\right), \\
&=&m_0\left(\begin{array}{cc}1/\sqrt{2}&1/\sqrt{2}\\-1/\sqrt{2}&1/\sqrt{2}\end{array}\right)\left(\begin{array}{cc}1&\epsilon\\-\epsilon&1\end{array}\right)
\left(\begin{array}{cc}-1+\alpha&0\\0&1+\beta\end{array}\right)
 \left(\begin{array}{cc}1&-\epsilon\\ \epsilon&1\end{array}\right)\left(\begin{array}{cc}1/\sqrt{2}&-1/\sqrt{2}\\1/\sqrt{2}&1/\sqrt{2}\end{array}\right)+O\left(\frac{m_1^2}{m_0^2}\right). \label{M}
\end{eqnarray} 
Here, $\alpha,\beta=\frac{m_1}{2m_0}(a+b\mp 2c)$ and $\epsilon=\frac{m_1}{4m_0}(a-b)$. From Eq.~(\ref{M}) it is easy to read off the mass-squared splitting induced by the perturbation $M_1$, 
\begin{equation}
\frac{\Delta m^2}{m_0^2}=\left|(-1+\alpha)^2-(1+\beta)^2\right|\simeq2\left|\alpha+\beta\right|=2\frac{m_1}{m_0}\left|a+b\right|,
\label{deltam_sol1}
\end{equation} 
 (higher order terms have been dropped) and the leptonic mixing matrix,
 \begin{equation}
 U_{MNS}=\frac{1}{\sqrt{2}}\left(\begin{array}{cc}1-\epsilon&1+\epsilon\\-1-\epsilon&1-\epsilon\end{array}\right).
 \end{equation}
One way to quantify how far from maximal is the mixing angle is to compute $|\cos2\theta|=|\cos^2\theta-\sin^2\theta|$. From the expression above, ignoring higher order terms,
\begin{equation}
|\cos2\theta|=\frac{1}{2}\left|(1-\epsilon)^2-(1+\epsilon)^2\right|=2\left|\epsilon\right|=\frac{m_1}{2m_0}\left|a-b\right|=\frac{\Delta m^2}{4m_0^2}\frac{\left|a-b\right|}{\left|a+b\right|}.
\end{equation}
What do these results mean? If these are compared to the experimental data, it is natural to choose $m_0^2$ to be the atmospheric mass-squared, $\Delta m^2_{\rm atm}\simeq 2\times10^{-3}$~eV$^2$  (see \cite{osc_anal} for a recent three-flavor analysis of the combined solar, atmospheric, and KamLAND data) and to associate the induced $\Delta m^2$, Eq.~(\ref{deltam_sol1}) to $\Delta m^2_{\rm sol}\simeq 7\times 10^{-5}$~eV$^2$, and $\theta$ to the atmospheric angle $\sin^2\theta_{23}\simeq0.5$. Hence, assuming $|a+b|,|a-b|$ both to be order one,
\begin{eqnarray}
m_0\simeq 4\times 10^{-2}, \\
m_1\simeq 9\times 10^{-4}, \\
|\cos2\theta_{23}|\lesssim0.01. \label{cos2th_21}
\end{eqnarray}
 
 Next, consider the so-called ``democratic" texture, namely
 \begin{equation}
 M_0=\frac{m_0}{2}\left(\begin{array}{cc}
 1&1\\1&1
\end{array}\right), 
\label{democratic}
 \end{equation}
with eigenvalues $m_0,0$ and eigenvectors $(1/\sqrt{2},1/\sqrt{2})^{\top}$ and $(1/\sqrt{2},-1/\sqrt{2})^{\top}$, respectively. In this case, Eq.~(\ref{democratic})+ Eq.~(\ref{M1_2}) is, ignoring higher order terms (this is very similar to the previous case, Eq.~(\ref{M})),
\begin{equation}
M_0+M_1=m_0\left(\begin{array}{cc}1/\sqrt{2}&1/\sqrt{2}\\-1/\sqrt{2}&1/\sqrt{2}\end{array}\right)\left(\begin{array}{cc}1&\epsilon\\-\epsilon&1\end{array}\right)
\left(\begin{array}{cc}\alpha&0\\0&1+\beta\end{array}\right)
 \left(\begin{array}{cc}1&-\epsilon\\ \epsilon&1\end{array}\right)\left(\begin{array}{cc}1/\sqrt{2}&-1/\sqrt{2}\\1/\sqrt{2}&1/\sqrt{2}\end{array}\right)+O\left(\frac{m_1^2}{m_0^2}\right),
\end{equation}
where $\alpha,\beta=\frac{m_1}{2m_0}(a+b\mp 2c)$, $\epsilon=\frac{m_1}{2m_0}(a-b)$. The interpretation of the different terms in this case is somewhat  different. As before, one can naturally associate $m_0^2$ to $\Delta m^2_{\rm atm}$ while it is natural to associate the shift of the zero eigenvalue of Eq.~(\ref{democratic}) to $\sqrt{\Delta m^2_{\rm sol}}$: 
\begin{equation}
\frac{\Delta m^2_{\rm sol}}{\Delta m^2_{\rm atm}}\simeq \alpha^2=\frac{m_1^2}{4m_0^2}\left(a+b-2c\right)^2. 
\end{equation} 
The modified atmospheric angle is, hence,
\begin{equation}
|\cos2\theta_{23}|=\frac{m_1}{m_0}\left|a-b\right|\simeq\sqrt{\frac{\Delta m^2_{\rm sol}}{\Delta m^2_{\rm atm}}}
\frac{2\left|a-b\right|}{\left|a+b-2c\right|}\lesssim 0.2, \label{cos2th_22}
\end{equation}
where, in the last step, $|a-b|,(a+b-2c)^2$ where replaced by 1 and  $3\pi/2$, respectively, their average values.\footnote{See the Appendix for the prescription for combining order one ``random" factors.}

Eqs.~(\ref{cos2th_21},\ref{cos2th_22}) will be considered as the upper bounds on $|\cos2\theta_{23}|$ for the two different ``models.'' Note that while the two $M_0$ yielded the same zeroth order physics, Eq.~(\ref{cos2th_21}) and Eq.~(\ref{cos2th_22}) differ by an order of magnitude! The reason for this is simple: in the case of  Eq.~(\ref{cos2th_22}), the perturbation parameter $m_0/m_1$ was chosen much larger than in the case of Eq.~(\ref{cos2th_21}) in order to fulfill the same intuitive agreement with the ``sub-leading" oscillation parameters. 
It will become clear in the next section that this discussion of $2\times 2$ toy models does indeed capture the relevant features of more realistic $3\times 3$ mass matrices. 

\setcounter{footnote}{0}
\setcounter{equation}{0}
\section{Case Studies}
\label{3}

In this section, several examples of $3\times3$ mass-matrices $M_0+M_1$ that yield, at zeroth order, maximal atmospheric mixing are inspected. Some of these have already been identified in previous studies, and have been used to determine, for example, expected values of $|U_{e3}|$ for different neutrino mass textures \cite{models}. Different $M_0$ differ by the values of other parameters, both known and unknown, such as the solar mixing angle and the hierarchy of the neutrino masses. The approach used in the previous section will be employed in order to estimate the maximal $|\cos2\theta_{23}|$ that each $M_0$ allows given a random perturbation $M_1$. The philosophy of studying the effect of random, sub-leading perturbations on flavor structures was already followed, for example, in \cite{aguilar_saavedra}.
 
Before proceeding, it is worthwhile to comment on $|\cos2\theta_{23}|$ as the relevant measure of deviation-from-maximal. Maximal atmospheric mixing means $|U_{\mu3}|=1/2$, such that deviation from maximal is proportional to $|U_{\mu3}|^2-1/2=-1/2\cos2\theta_{23}-\sin^2\theta_{13}\sin^2\theta_{23}$.  Note that the second term in the previous expression is proportional to $|U_{e3}|^2$, much smaller than the first term (this can be easily verified {\it a posteriori} for all mass-matrices discussed here), such that $|U_{\mu3}|^2-1/2\simeq-1/2\cos2\theta_{23}$. Hence, $\cos2\theta_{23}$ is a {\it bona fide}\/ measure of deviation-from-maximal-atmospheric-mixing.
 
 Analogous to the previous section (Eq.~(\ref{M1_2})),
 \begin{equation}
 M_1\equiv m_1\left(\begin{array}{ccc}a&d&e\\d&b&f\\e&f&c\end{array}\right), \label{M1_3}
 \end{equation}
will be assumed for all the discussions that follow. $a,b,c,d,e,f$ are random order one real numbers.

\subsection{Normal Hierarchy} 

A ``normal" mass-hierarchy is characterized by $\Delta m^2_{13}>0$, where $m_3^2\gg m_1^2,m_2^2$. There are several mass-matrix textures that lead to two massless and one massive eigenstate, the latter composed of a 50--50 $\nu_{\mu}-\nu_{\tau}$ mixture. One possibility is to introduce one right-handed neutrino with a very heavy Majorana mass, coupled with identical strength to ${\rm L}_{\mu}$ and ${\rm L}_{\tau}$\footnote{${\rm L}_{\ell}$ are the left-handed lepton weak doublets, $(\nu_{\ell},\ell)^{\top}$.}  plus the Higgs doublet, in the basis where the charged lepton Yukawa couplings and the charged weak gauge couplings are diagonal \cite{srhnd,minimal_model}. Upon integrating out the right-handed fermion, and below the electroweak symmetry breaking scale,  
\begin{equation}
M_0=\frac{m_0}{2}\left(\begin{array}{ccc}0&0&0\\0&1&1\\0&1&1\end{array}\right)=m_0
\left(\begin{array}{ccc}1&0&0\\0&1/\sqrt{2}&1/\sqrt{2}\\0&-1/\sqrt{2}&1/\sqrt{2}\end{array}\right)
\left(\begin{array}{ccc}0&0&0\\0&0&0\\0&0&1\end{array}\right)
\left(\begin{array}{ccc}1&0&0\\0&1/\sqrt{2}&-1/\sqrt{2}\\0&1/\sqrt{2}&1/\sqrt{2}\end{array}\right)
\label{M0_3normal}
\end{equation}
is generated. Defining 
\begin{equation}
U_0\equiv\left(\begin{array}{ccc}1&0&0\\0&1/\sqrt{2}&1/\sqrt{2}\\0&-1/\sqrt{2}&1/\sqrt{2}\end{array}\right),
\end{equation}
$M_0+M_1$ can be written as
\begin{equation}
M_0+M_1=m_0U_0\left(\begin{array}{ccc}\alpha&\epsilon&\delta\\ \epsilon&\beta&\eta\\\delta&\eta&1+\gamma\end{array}\right)U_0^{\top},
\end{equation}
where
\begin{eqnarray}
&\alpha=\frac{m_1}{m_0}a,~~\beta=\frac{m_1}{m_0}\left(\frac{b+c}{2}-f\right),~~\gamma=\frac{m_1}{m_0}\left(\frac{b+c}{2}+f\right), \nonumber \\
&\epsilon=\frac{m_1}{m_0}\frac{d-e}{\sqrt{2}},~~\delta=\frac{m_1}{m_0}\frac{d+e}{\sqrt{2}},~~\eta=\frac{m_1}{m_0}\frac{b-c}{2}. \label{abcde}
\end{eqnarray}
Therefore, at leading order in $m_1/m_0$,
\begin{equation}
M_0+M_1=m_0U_0
\left(\begin{array}{ccc}\cos\omega&\sin\omega&\delta\\-\sin\omega&\cos\omega&\eta\\-\delta^{\prime}&-\eta^{\prime}&1\end{array}\right)
\left(\begin{array}{ccc}\alpha^{\prime}&0&0\\ 0&\beta^{\prime}&0\\ 0 & 0 &1+\gamma\end{array}\right)
\left(\begin{array}{ccc}\cos\omega&-\sin\omega&-\delta^{\prime}\\\sin\omega&\cos\omega&-\eta^{\prime}\\\delta&\eta&1\end{array}\right)U_0^{\top}.
\end{equation}
$\alpha^{\prime}$ and $\beta^{\prime}$ are the eigenvalues (ordered in ascending order of absolute value) and $\omega$ the mixing angle of the $2\times 2$ sub-matrix $\left(\begin{array}{cc}
 \alpha&\epsilon\\\epsilon&\beta\end{array}\right)$, while $\delta^{\prime}=\delta\cos\omega-\eta\sin\omega$ and $\eta^{\prime}=\delta\sin\omega+\eta\cos\omega$. The two independent mass-squared differences are:
 \begin{equation}
 \Delta m^2_{13}=m_0^2(1+2\gamma),~~~~\frac{\Delta m^2_{12}}{\Delta m^2_{13}}=\beta^{\prime2}-\alpha^{\prime2},
 \end{equation}
while the mixing matrix is\footnote{The PDG \cite{PDG} definition of the leptonic mixing matrix is $\nu_{\alpha}=U_{\alpha\beta}\nu_i$ for
$\alpha=e,\mu,\tau$ and $i=1,2,3$, and the mass matrix in the flavor basis is given by $U^*{\rm diagonal}[m_1,m_2,m_3]U^{\dagger}$. Hence, Eq.~(\ref{MNSnormal}) is equal to $U^*$. However, since I am restricted to real mass matrices, $U=U^*$ and this distinction can be safely ignored.} 
\begin{equation}
U_{MNS}=\left(\begin{array}{ccc}
\cos\omega&\sin\omega&\delta \\
\star&\star&1/\sqrt{2}(1+\eta) \\
\star&\star&1/\sqrt{2}(1-\eta)
\end{array}\right), \label{MNSnormal}
\end{equation}
where only the relevant elements have been spelled out (others have been mercifully represented by $\star$). One can then ``read off"
\begin{eqnarray}
&\theta_{12}=\omega,~~~~~|U_{e3}|=|\delta|=\frac{m_1}{m_0}\frac{\left|d+e\right|}{2}, \\
&|\cos2\theta_{23}|=\frac{1}{2}|\left(1+\eta)^2-(1-\eta^2)\right|=2|\eta|=\frac{m_1}{m_0}\left|b-c\right|.
\end{eqnarray}
As done in the previous section, estimates for the sizes of various parameters are obtained after $m_1/m_0$ is extracted from the fact that the ratio of mass-squared differences is known: $\Delta m^2_{12}/\Delta m^2_{13}\simeq 0.034$. More quantitatively, it has been estimated that, at the three sigma confidence level, $0.018<\Delta m^2_{12}/\Delta m^2_{13}<0.054$ \cite{osc_anal}. A revised analyses of the Super-Kamiokande atmospheric data, however, seems to point to a slightly smaller value for the atmospheric mass-squared difference (see \cite{SuperK_new}. There is, currently, no publication of this reanalysis). In order to take this into account, I simply rescale the results of \cite{osc_anal} by the ratio of the ``old" and ``new" central values $(2.6\times 10^{-3}/2.0\times 10^{-3})$:
\begin{equation}
0.023<\frac{\Delta m^2_{12}}{|\Delta m^2_{13}|}<0.070,
\label{range}
\end{equation} 
which should be very close to the real three sigma confidence level range. For all estimates bellow, in order to be conservative, I'll use the upper bound of Eq.~(\ref{range}), which yields the loosest constraints on $\cos2\theta_{23}$.
In the case at hand
\begin{equation}
\frac{\Delta m^2_{12}}{\Delta m^2_{13}}=\left|\beta+\alpha\right|\sqrt{\left(\beta-\alpha\right)^2+4\epsilon^2}=\frac{m_1^2}{m^2_0}\left|a+\frac{b+c}{2}-f\right|\sqrt{\left(a-\frac{b+c}{2}+f\right)^2+2(d-e)^2}.
\end{equation}
Replacing the quantity inside the absolute value by its average value (see Appendix), 
\begin{equation}
\frac{m_1^2}{m_0^2}\simeq\frac{0.070}{1.5}\simeq(0.2)^2,
\end{equation}
such that 
\begin{equation}
|U_{e3}|\lesssim\frac{0.2}{2}\sqrt{\frac{2}{3}}\simeq0.09, ~~~~~|\cos2\theta_{23}|\lesssim 0.2\sqrt{\frac{2}{3}}\simeq0.18.
\end{equation}
In summary, it is expected that the solar angle is ``anarchical," {\it i.e.}, dictated by the diagonalization of a $2\times 2$ sub-matrix with random order one entries. This hypothesis fits the data very well \cite{anarchy_stat}. On the other hand, both $|U_{e3}|$ and the deviation of $\theta_{23}$ from maximal are of order $\sqrt{\Delta m^2_{12}/\Delta m^2_{13}}$, $O(10^{-1})$. 

A couple of important comments are warranted. The average {\sl upper bound} on $|U_{e3}|^2\lesssim0.008$ is smaller than the estimated {\sl lower bound} on $|U_{e3}|^2$ obtained from the anarchical hypothesis \cite{anarchy_stat} (albeit only slightly), meaning that next-generation experiments sensitive to $|U_{e3}|^2\gtrsim 10^{-2}$ may be able to start distinguishing the two paradigms. The average deviation of $\theta_{23}$ from maximal is rather large, but well within current bounds \cite{osc_anal}:
\begin{equation}
|\cos2\theta_{23}|<0.34~(0.44),~{\rm at}~2\sigma~(3\sigma).
\label{bound_23}
\end{equation}

Finally, it should be noted that given the approach adopted here, nothing can be said about the sign of $\cos2\theta_{23}$, {\it i.e.}, whether the $\nu_3$ state is ``predominantly $\nu_{\mu}$'' ($\cos2\theta_{23}<0$) or ``predominantly $\nu_{\tau}$'' ($\cos2\theta_{23}>0$). Note that we already know, experimentally, that $\nu_3$ is not ``predominantly $\nu_e$" ($|U_{e3}|^2<0.5$) and that, thanks to matter effects in the Sun's core, $\nu_1$ is ``predominantly $\nu_e$" ($\cos2\theta_{12}>0$, the ``light-side" of the solar neutrino parameter space \cite{dark_side}) at the five sigma confidence level \cite{osc_anal}.

\subsection{Inverted Hierarchy}
An ``inverted'' mass-hierarchy is characterized by $\Delta m^2_{13}<0$, and two eigenstates that are almost degenerate in mass-squared. As in the normal hierarchy case, there are several textures that yield two massive mass-eigenstates, degenerate in mass-squared, and a  massless one, composed of a 50--50 $\nu_{\mu}-\nu_{\tau}$ mixture. I'll look at three different cases, characterized by whether the massive mass-eigenstates are degenerate in mass (as opposed to mass-squared) or by how the massive mass-eigenstates ``share" the $\nu_e$ content.

The case which is ``most similar" to the one in the previous subsection is well represented by 
\begin{equation}
M_0=m_0\left(\begin{array}{ccc}1&0&0\\0&1/2&-1/2\\0&-1/2&1/2\end{array}\right)=
m_0U_0
\left(\begin{array}{ccc}1&0&0\\0&1&0\\0&0&0\end{array}\right)
U_0^{\top}.
\end{equation}
Note that the heavy mass-eigenstates are degenerate in mass. $M_0+M_1$ can be written as
\begin{equation}
M_0+M_1=m_0U_0\left(\begin{array}{ccc}1+\alpha&\epsilon&\delta\\ \epsilon&1+\beta&\eta\\\delta&\eta&\gamma\end{array}\right)U_0^{\top},
\end{equation}
where $\alpha$, $\beta$, $\gamma$, $\delta$, $\epsilon$, $\delta$, and $\eta$ are as in Eq.~(\ref{abcde}).

As in the previous case, degenerate perturbation theory yields
\begin{equation}
M_0+M_1=m_0U_0
\left(\begin{array}{ccc}\cos\omega&\sin\omega&-\delta\\-\sin\omega&\cos\omega&-\eta\\\delta^{\prime}&\eta^{\prime}&1\end{array}\right)
\left(\begin{array}{ccc}1+\alpha^{\prime}&0&0\\ 0&1+\beta^{\prime}&0\\ 0 & 0 &\gamma\end{array}\right)
\left(\begin{array}{ccc}\cos\omega&-\sin\omega&\delta^{\prime}\\\sin\omega&\cos\omega&\eta^{\prime}\\-\delta&-\eta&1\end{array}\right)U_0^{\top},
\end{equation}
where the primed quantities and $\omega$ are defined as in the previous subsection.

The extraction of oscillation parameters proceeds as before. Indeed, the MNS matrix is identical to Eq.~(\ref{MNSnormal}), up to $\delta,\eta\to-\delta,-\eta$, while the mass-squared differences are 
\begin{equation}
\Delta m^2_{13}=-m_0^2,~~~~\frac{\Delta m^2_{12}}{|\Delta m^2_{13}|}=2\left(\beta^{\prime}-\alpha^{\prime}\right).
\end{equation}
Note that $\Delta m^2_{13}$ also receives a higher order ($O(m_0m_1)$) correction, which can be safely neglected. Furthermore, As before, we extract $m_1/m_0$ by requiring
\begin{equation}
\frac{\Delta m^2_{12}}{\Delta m^2_{13}}=2\sqrt{\left(\beta-\alpha\right)^2+4\epsilon^2}=2\frac{m_1}{m_0}\sqrt{\left(a-\frac{b+c}{2}+f\right)^2+2(d-e)^2}\simeq0.07,
\end{equation}
where I conservatively use the upper bound of Eq.~(\ref{range}). One obtains, on average, $m_1/m_0\simeq0.02$, which implies 
\begin{equation}
|U_{e3}|\lesssim\frac{0.02}{2}\sqrt{\frac{2}{3}}\simeq0.008, ~~~~~|\cos2\theta_{23}|\lesssim0.02\sqrt{\frac{2}{3}}\simeq0.016.
\end{equation}
Hence, in the case of an inverted mass-hierarchy with two mass-degenerate (at leading order) eigenstates, one expects a large, ``random" solar angle, and very small values for $|U_{e3}|$ and $\cos2\theta_{23}$, both of order $\Delta m^2_{12}/|\Delta m^2_{13}|$ ($O(10^{-2})$), well within current experimental bounds.

\subsection{Inverted Hierarchy, Traceless}
There is the possibility that the two massive zeroth-order mass-eigenstates have opposite CP-parities ({\it i.e.,}\/ $m_{0,1}=-m_{0,2}$). One example is
\begin{equation}
M_0=m_0\left(\begin{array}{ccc}1&0&0\\0&-1/2&1/2\\0&1/2&-1/2\end{array}\right)=
m_0U_0
\left(\begin{array}{ccc}1&0&0\\0&-1&0\\0&0&0\end{array}\right)
U_0^{\top}. \label{M0_3it}
\end{equation}
Similar to the previous two cases, $M_0+M_1$ can be written as
\begin{equation}
M_0+M_1=m_0U_0\left(\begin{array}{ccc}1+\alpha&\epsilon&\delta\\ \epsilon&-1+\beta&\eta\\\delta&\eta&\gamma\end{array}\right)U_0^{\top},
\end{equation}
where $\alpha$, $\beta$, $\gamma$, $\delta$, $\epsilon$, $\delta$, and $\eta$ are as in Eq.~(\ref{abcde}).

In this case, the diagonalization procedure is somewhat different, as there are no degenerate eigenvalues (it is, indeed, simpler). To lowest order in $m_1/m_0$,
\begin{equation}
M_0+M_1=m_0U_0
\left(\begin{array}{ccc}1&-\epsilon/2&-\delta\\\epsilon/2&1&\eta\\\delta&-\eta&1\end{array}\right)
\left(\begin{array}{ccc}1+\alpha&0&0\\ 0&-1+\beta&0\\ 0 & 0 &\gamma\end{array}\right)
\left(\begin{array}{ccc}1&\epsilon/2&\delta\\-\epsilon/2&1&-\eta\\-\delta&\eta&1\end{array}\right)U_0^{\top}.
\end{equation}
The two independent mass-squared differences are, to lowest order:
 \begin{equation}
 \Delta m^2_{13}=-m_0^2,~~~~\frac{\Delta m^2_{12}}{|\Delta m^2_{13}|}=2|\beta+\alpha|=\frac{m_1}{m_0}|2a+b+c-2f|,
 \end{equation}
while the mixing matrix is
\begin{equation}
U_{MNS}=\left(\begin{array}{ccc}
1&-\epsilon/2&-\delta \\
\star&\star&1/\sqrt{2}(1+\eta) \\
\star&\star&1/\sqrt{2}(1-\eta)
\end{array}\right). 
\end{equation}
Hence
\begin{eqnarray}
&|\sin\theta_{12}|=\frac{\epsilon}{2}=\frac{m_1}{2\sqrt{2}m_0}|d-e|,~~~~~|U_{e3}|=|\delta|=\frac{m_1}{m_0}\frac{\left|d+e\right|}{2}, \\
&|\cos2\theta_{23}|=\frac{1}{2}|\left(1+\eta)^2-(1-\eta^2)\right|=2|\eta|=\frac{m_1}{m_0}\left|b-c\right|.
\end{eqnarray}
Following the by-now-familiar procedure, one estimates 
\begin{equation}
|\sin\theta_{12}|\lesssim\frac{\Delta m^2_{12}}{|\Delta m^2_{13}|}\frac{\sqrt{2}}{2\sqrt{5}},~~~~~|U_{e3}|\lesssim\frac{\Delta m^2_{12}}{|\Delta m^2_{13}|}\frac{1}{2\sqrt{5}},~~~~~ 
|\cos2\theta_{23}|\lesssim\frac{\Delta m^2_{12}}{|\Delta m^2_{13}|}\frac{1}{\sqrt{5}}.
\end{equation}
Note that, here, not only are $|U_{e3}|$ and $\cos2\theta_{23}$ smaller than $\Delta m^2_{12}/|\Delta m^2_{13}|$, but so is the solar angle (indeed, it is, {\sl  on average}, smaller than $\theta_{13}$!). Given the current lower
bound on the solar mixing angle, $\sin^2\theta_{12}>0.22$ at the three sigma confidence level \cite{osc_anal}, one can only conclude that such a scenario is experimentally ruled out, as the estimate above is $\sin^2\theta_{12}\lesssim0.0005$.  

One should not, however, conclude that the mass-matrix texture Eq.~(\ref{M0_3it}) is currently ruled out, but that if it is to work, it has to be ``perturbed" by a rather ``structured" matrix in order to guarantee that the solar angle is very strongly modified from its leading order value, while, say, $|U_{e3}|$ is not. This could be accomplished, for example, if $d$ is close to $(-e)$ and much larger than $a,b,c,f$.\footnote{This should be done with care. $M_1$ stops being a perturbation as $d$ approaches $m_0/m_1$, indicating i-that the approximate results obtained here need not be reliable, ii-that the zeroth-order mass-matrix texture has to be revised in order to somehow include the information regarding the large-but-not-maximal solar angle. Such examples can be found, for example in \cite{tri-bimaximal}, and references therein.} 

\subsection{Inverted Hierarchy, Traceless, Bi-Maximal Mixing}
An interesting mass-matrix texture is motivated by an $L_e-L_{\mu}-L_{\tau}$ discrete symmetry. It is 
\begin{equation}
M_0=\frac{m_0}{\sqrt{2}}\left(\begin{array}{ccc}0&1&1\\1&0&0\\1&0&0\end{array}\right)=m_0
\left(\begin{array}{ccc}1/\sqrt{2}&1/\sqrt{2}&0\\1/2&-1/2&-1/\sqrt{2}\\1/2&-1/2&1/\sqrt{2}\end{array}\right)
\left(\begin{array}{ccc}1&0&0\\0&-1&0\\0&0&0\end{array}\right)
\left(\begin{array}{ccc}1/\sqrt{2}&1/2&1/2\\1/\sqrt{2}&-1/2&-1/2\\0&-1/\sqrt{2}&1/\sqrt{2}\end{array}\right). \label{M0_3ibm}
\end{equation}
It proves economical to define 
\begin{equation}
U_0^{bm}\equiv\left(\begin{array}{ccc}1/\sqrt{2}&1/\sqrt{2}&0\\1/2&-1/2&-1/\sqrt{2}\\1/2&-1/2&1/\sqrt{2}\end{array}\right). \label{Ubm}
\end{equation}
Note that not only do the massive mass-eigenstates have opposite CP-parity, but they also ``share" the $\nu_e$ content equally ($|U^{bm}_{0,e1}|^2=|U^{bm}_{0,e2}|^2=1/2$). Eq.~(\ref{Ubm}) is often referred to as a bi-maximal mixing matrix \cite{bi-maximal} ($\sin^2\theta_{23}=\sin^2\theta_{12}=1/2$, $\sin^2\theta_{13}=0$), and has received a significant amount of attention in the literature.
\begin{equation}
M_0+M_1=m_0U_0^{bm}\left(\begin{array}{ccc}1+\alpha&\epsilon&\delta\\ \epsilon&-1+\beta&\eta\\\delta&\eta&\gamma\end{array}\right)U_0^{bm\top},
\end{equation}
where
\begin{eqnarray}
&\alpha=\frac{m_1}{4m_0}\left(2a+b+c+2\sqrt{2}(d+e)+2f\right),~~\beta=\frac{m_1}{4m_0}\left(2a+b+c-2\sqrt{2}(d+e)+2f\right),~~\gamma=\frac{m_1}{m_0}\left(\frac{b+c}{2}-f\right), \nonumber \\
&\epsilon=\frac{m_1}{4m_0}\left(2a-b-c-2f\right),~~\delta=\frac{m_1}{4m_0}\left(-\sqrt{2}(b-c)-2d+2e\right),~~\eta=\frac{m_1}{4m_0}\left(\sqrt{2}(b-c)-2d+2e\right). 
\end{eqnarray}

Identical to the previous subsection, to lowest order in $m_1/m_0$,
\begin{equation}
M_0+M_1=m_0U_0^{bm}
\left(\begin{array}{ccc}1&-\epsilon/2&-\delta\\\epsilon/2&1&\eta\\\delta&-\eta&1\end{array}\right)
\left(\begin{array}{ccc}1+\alpha&0&0\\ 0&-1+\beta&0\\ 0 & 0 &\gamma\end{array}\right)
\left(\begin{array}{ccc}1&\epsilon/2&\delta\\-\epsilon/2&1&-\eta\\-\delta&\eta&1\end{array}\right)U_0^{bm\top},
\end{equation}
and the two independent mass-squared differences are, at leading order:
 \begin{equation}
 \Delta m^2_{13}=-m_0^2,~~~~\frac{\Delta m^2_{12}}{|\Delta m^2_{13}|}=2|\beta+\alpha|=\frac{m_1}{2m_0}|2a+b+c+2f|.
 \end{equation}
The leptonic mixing matrix is
\begin{equation}
U_{MNS}=\left(\begin{array}{ccc}
1/\sqrt{2}(1+\epsilon/2)&1/\sqrt{2}(1-\epsilon/2)&1/\sqrt{2}(\eta-\delta) \\
\star&\star&1/\sqrt{2}(1+1/\sqrt{2}(\eta+\delta)) \\
\star&\star&1/\sqrt{2}(1-1/\sqrt{2}(\eta+\delta))
\end{array}\right), \label{MNSbimaximal}
\end{equation}
such that 
\begin{equation}
|\cos2\theta_{12}|=\epsilon=\frac{m_1}{4m_0}|2a-b-c-2f|,~~~|\cos2\theta_{23}|=\sqrt{2}|\delta+\eta|=\frac{\sqrt{2}m_1}{m_0}|e-d|,~~~|U_{e3}|=\frac{\eta-\delta}{\sqrt{2}}=\frac{m_1}{2m_0}|b-c|.
\end{equation}
Solving for $m_1/m_0$ in terms of the ratio of the mass-squared differences and replacing the order one expressions by their average values
\begin{equation}
|\cos2\theta_{12}|\lesssim\frac{\Delta m^2_{12}}{2|\Delta m^2_{13}|},~~~|\cos2\theta_{23}|\lesssim\frac{\Delta m^2_{12}}{|\Delta m^2_{13}|}\frac{2\sqrt{2}}{\sqrt{5}},~~~|U_{e3}|\lesssim\frac{\Delta m^2_{12}}{|\Delta m^2_{13}|}\frac{1}{\sqrt{5}}.
\end{equation}
Similar to the previous subsection, this scenario runs afoul of the experimental data, because it predicts an incorrect value for the solar angle. It requires, on average, that solar mixing is too close to maximal: $\cos2\theta_{12}\lesssim 0.04$, while current data indicate $0.22<\cos2\theta_{12}<0.56$ at the three sigma level \cite{osc_anal}. As in the case of the previous subsection, this result does not imply that the texture Eq.~(\ref{M0_3ibm}) is ruled out, but simply that the ``perturbation" has to be such that the solar angle is significantly modified from its zeroth order value, while other parameters are less ``perturbed." This could be accomplished, for example, if $a$ is of order $(-f)$, and much larger than $b,c,d,e$ (but see footnote in subsection \ref{3}C).  

There are other zeroth order bi-maximal textures that do not suffer from a too-large $\theta_{12}$. The mass-matrix
\begin{equation}
M_0=m_0\left(\begin{array}{ccc}1&0&0\\0&1/2&1/2\\0&1/2&1/2\end{array}\right)=
m_0U_0^{bm}
\left(\begin{array}{ccc}1&0&0\\0&1&0\\0&0&0\end{array}\right)
U_0^{bm\top}.
\end{equation}
has two degenerate eigenvalues, indicating that the zeroth order solar angle is very sensitive to a generic perturbation, no matter how weak (this was observed, for example, in subsection \ref{3}B). In this case,  one expects that both $|U_{e3}|$ and $|\cos2\theta_{23}|$ are small, of order $\Delta m^2_{12}/|\Delta m^2_{13}|$, while the solar angle is ``anarchical."

Finally, I'll conclude with a few words on textures that yield three mass-squared-degenerate eigenstates, some of which can be found in \cite{models}. In this case, it turns out that it is possible to obtain, for example, an anarchical solar angle while maintaining a generically small $\cos2\theta_{23}$ and $|U_{e3}|$. The biggest challenge, one that cannot be overcome by a generic perturbation of the type Eq.~(\ref{M1_3}), is generating simultaneously the two distinct mass-squared differences. This is very simple to understand. The three eigenvalues which are degenerate at zeroth order will be modified to 
$m_i=m_0(\pm1+\zeta_i)$, where $\zeta_i$ are order $m_1/m_0$. Hence, $\Delta m^2_{12}\simeq2(\zeta_2-\zeta_1)$, 
 $\Delta m^2_{13}\simeq2(\zeta_3-\zeta_1)$, which leads to $\Delta m^2_{12}\simeq\Delta m^2_{13}$, in conflict with the experimental data. Note that in the case of degenerate zeroth order eigenvalues it is quite unlikely that both mass-squared differences are induced by a single random perturbation -- significantly less likely than in the original anarchy proposal \cite{anarchy}. In an ``anarchical model'' that claims to explain both the mixing matrix and the mass-eigenvalues as typical fluctuations of a random mass-matrix (see \cite{anarchy}) the requirement on the eigenvalues of the mixing matrix is that they are ``separated" by a factor $\sqrt{|\Delta m^2_{13}|/{\Delta m^2_{12}}}\simeq 5$. In zeroth-order-degenerate models, the requirement imposed on the eigenvalues of the random perturbation-matrix is that their ratio is of order $|\Delta m^2_{13}|/\Delta m^2_{12}\simeq30$.
 
 As before, degenerate mass-matrix textures are certainly not ruled out, but they require a somewhat ``structured perturbation" (see \cite{a4} for an example) in order to properly accommodate the neutrino data. In some cases, the leptonic mixing matrix can still turn out to be completely anarchical.  

\section{Discussion, Summary, and Conclusions}

The fact that the atmospheric mixing angle is currently consistent with maximal is often interpreted as evidence that there is a fundamental symmetry or dynamical reason behind the observed values of lepton masses and mixing. Indeed, several flavor models have been explored in the past that yield, at leading order, maximal atmospheric mixing and zero $U_{e3}$. Sub-leading effects, however, will almost always lead to departure from these leading order predictions. These include spontaneous or explicit symmetry breaking effects and renormalization group running, and are often responsible for explaining the ``solar" part of the neutrino oscillation parameters.

It is important to understand how close to maximal such models predict the atmospheric mixing angle to be in order to determine whether the current or next generation of neutrino oscillation experiments is capable of favoring or disfavoring this particular approach to ``explaining" the leptonic sector. Currently, the atmospheric angle is only loosely constrained by atmospheric neutrino data: $-0.44<\cos2\theta_{23}<0.44$ at the three sigma confidence level (Eq.~(\ref{bound_23})).\footnote{It is curious that the measurement of the atmospheric angle is significantly less precise than the measurement of the solar angle. Currently, $0.22<\cos2\theta_{12}<0.56$ at the three sigma level. One of the reasons for this is the fact that strong solar matter effects render the survival probability of electron-type solar neutrinos proportional to $\sin^2\theta_{12}$ for a large range of solar neutrino energies, and hence the solar data is very sensitive to $\theta_{12}$.}

The current generation of long-baseline accelerator experiments is not expected to improve significantly on the current bound \cite{MINOSetal}, while studies of next-generation ``off-axis" experiments have claimed that these are able to measure $\sin^22\theta_{23}$ at the 1\% level \cite{superbeam}. More detailed studies, however, are still absent \cite{bernstein}. Note that while a 1\% measurement of $\sin^22\theta_{23}$ sounds impressive, for a best-fit point $\sin^22\theta_{23}=1$ it ``only" translates into $|\cos2\theta_{23}|<0.1$. Other constraint may be provided by future atmospheric neutrino studies \cite{atmospheric_new,peres_smirnov}, or, ultimately, by neutrino factories, which may be able to measure $\cos2\theta_{23}$ at the few percent level \cite{nufact}.

I have explored different neutrino Majorana mass-matrices that yield, at zeroth order, a $\nu_3$ state which is $0\%$ $\nu_e$,
$50\%$ $\nu_{\mu}$, and $50\%$ $\nu_{\tau}$ and are ``perturbed" by a random Majorana mass-matrix. The magnitude of the perturbation is chosen such that one can also account for the solar mass-squared difference. In order to extract numerical results from the random mixing matrices, I computed the absolute values of different observables, and replaced the order one numbers by their average values. This procedure allows one to easily define the magnitude of the perturbation and estimate how much $|U_{\mu3}|$ ($|U_{e3}|$) is expected to deviate on average from $1/2$ (0). A summary of the results obtained is presented in Table~\ref{sumtable}.

For all mass-textures examined, the following features are observed. Both $|U_{e3}|$ and $|\cos2\theta_{23}|$ are, on average, of the same order of magnitude, as one should expect from a featureless perturbation (this is obviously not the case for all neutrino mass models. See, for example, \cite{Grimus_Lavoura} for a counter-example). Furthermore, they are of order $\Delta m^2_{12}/|\Delta m^2_{13}|$ for textures that yield an inverted mass-hierarchy or of order $\sqrt{\Delta m^2_{12}/|\Delta m^2_{13}|}$ for textures that yield a normal mass-hierarchy. As far as the solar mixing angle is concerned, it is either ``anarchical" (order one), in good agreement with the data (this is the case of mass-textures that yield $\nu_1$ and $\nu_2$ states that are mass-degenerate at zeroth order), or doomed to be, on average, way too small/too large, in which case, the current approach safely does not yield a good fit to the data. 

\renewcommand\arraystretch{2.0}
\begin{table}
\caption{Order of magnitude for the average value of the leptonic mixing parameters for the different neutrino mass-textures. See Sec.~\ref{3} for more detail. Also included are the expectations for an anarchical mixing matrix \cite{anarchy,anarchy_stat}.}
\label{sumtable}
\begin{tabular}{|c|c|c|c|} \hline
Texture  (Sec.) & $|U_{e3}|$ & $|\cos2\theta_{23}|$ & Solar Angle \\ \hline  
Normal Hierarchy (\ref{3}A) & $\sqrt{{\Delta m^2_{12}}/{\Delta m^2_{13}}}$ & $\sqrt{{\Delta m^2_{12}}/{\Delta m^2_{13}}}$ & O(1) \\ \hline
Inverted Hierarchy (\ref{3}B) & ${\Delta m^2_{12}}/{|\Delta m^2_{13}|}$ & ${\Delta m^2_{12}}/{|\Delta m^2_{13}|}$ & O(1) \\ \hline
Inverted Hierarchy, Traceless (\ref{3}C) &  ${\Delta m^2_{12}}/{|\Delta m^2_{13}|}$ & ${\Delta m^2_{12}}/{|\Delta m^2_{13}|}$ & 
$|\sin\theta_{12}|\sim{\Delta m^2_{12}}/{|\Delta m^2_{13}|}$ \\ \hline
Inverted Hierarchy, Bi-maximal \ref{3}(D) & ${\Delta m^2_{12}}/{|\Delta m^2_{13}|}$ & ${\Delta m^2_{12}}/{|\Delta m^2_{13}|}$ & 
$|\cos2\theta_{12}|\sim{\Delta m^2_{12}}/{|\Delta m^2_{13}|}$ \\ \hline
Anarchy & $>0.1$ & O(1) & O(1) \\ \hline
\end{tabular}
\end{table}

 What do these results mean? In the case of a normal mass-hierarchy, $|\cos2\theta_{23}|$ is, on average, significantly away from maximal, such that proposed off-axis experiments should observe a small but significant deviation of $\sin^22\theta_{23}$ from 1. Clearly, if they do not see such a deviation, it does not mean that mass textures that yield a normal mass hierarchy are ``ruled out" -- it simply means that they either ``got lucky" or that they are not ``perturbed by a random mass-matrix." 

 In the case of an inverted mass-hierarchy, $\theta_{23}$ is very close to maximal (and $|U_{e3}|^2$ is $\ll 10^{-2}$, a rather unfortunate case for the off-axis experiments under consideration and reactor searches for $|U_{e3}|$) and no deviation from maximal should be observed in next-generation experiments. Note that the average upper bounds calculated above can be violated by a factor of a few due to random statistical fluctuations of the order one parameters, but are probably not violated by more than an order of magnitude {\sl even for more structured models}. Keep in mind that the type of models under consideration here predict maximal mixing at zeroth order. Perturbations that are consistent with the solar data and that yield a much larger deviation of $\cos2\theta_{23}$ from zero will violate this assumption. Therefore, the experimental determination that $|\cos2\theta_{23}|$ is not consistent with zero at the $10\%$ level indicates that textures that yield maximal atmospheric mixing and an inverted mass-hierarchy are disfavored.
 
 Before concluding, I should mention that the analysis performed here is not, and is not meant to be, exhaustive. There are, potentially, other models that lead to maximal mixing in the atmospheric sector at zeroth order but are not well characterized by the mass-textures explored. I would also like to emphasize again that I only address textures that predict maximal atmospheric mixing at zeroth order. There are several models in the literature that ``predict," say, large atmospheric mixing and zero $U_{e3}$ at zeroth order (one simple example is the so-called ``semi-anarchical" texture of \cite{afm}). As far as these are concerned, the expected deviation of the atmospheric angle from $\pi/4$ is order 1. It should be noted that, according to the results on Sec.~\ref{3}A, it may not be easy to differentiate a texture like Eq.~(\ref{M0_3normal}) from the semi-anarchical one if $\cos2\theta_{23}$ is determined to be significantly away from 0 -- more detailed analyses would be required.
    
In conclusion, neutrino oscillations provide our first and, currently, only evidence for physics beyond the Standard Model. The amount of information available regarding the nature of this new physics is, however, very small. As far as the neutrino sector is concerned, and in the case of neutrino oscillation experiments in particular, it is possible that the best one can accomplish is to measure precisely the elements of the leptonic mixing matrix, the value of the mass-squared differences, and the neutrino mass-hierarchy. Hence, we need to take maximal advantage of all these measurements in order to come up with the most complete and useful description of Nature. 

I tried to point out in this exercise that the deviation of the atmospheric mixing angle from maximal is, potentially, an important observable, which may be able to test whether this maximal mixing phenomenon is somehow encoded in Nature via some fundamental principle, or whether it is simply accidental. It seems that even if maximal mixing is indeed fundamental at some energy scale, it is still true that, in the case of a normal mass hierarchy, it is possible that next-generation oscillation experiments will see a deviation of $\cos2\theta_{23}$ from 0, but more quantitative analysis of these experimental proposals are required in order to better quantify this statement. On the other hand, in the case of  and inverted mass-hierarchy, $\cos2\theta_{23}=0$ seems to be a robust prediction for any next-generation neutrino oscillation experiment, given that these probably cannot measure $\cos2\theta_{23}$ at better than 10\%.

{\bf NOTE ADDED:} After the completion of this manuscript, \cite{Petcov_new} became publicly available at the preprint archive. It contains a discussion of the effects of a ``perturbation" on a zeroth-order bi-maximal leptonic mixing matrix (the presentation and motivation are different from the one here. See \cite{Petcov_new} for details), and discusses relations between the atmospheric mixing angle and $|U_{e3}|$. It also contains some of the results presented in Sec.~\ref{3}D.   

\section*{Acknowledgments}

I am happy to thank Hitoshi Murayama for words of encouragement and Werner Rodejohann for useful comments. 

\appendix
\section{Combining Order One Random Entries}

In order to obtain numerical estimates for the products and sums of the random order one parameters $a,b,c,d,e,f$ defined in Eqs.~(\ref{M1_2},\ref{M1_3}), I follow the prescription below:

i-Assume that $a,b,c,d,e,f$ are uncorrelated random variables, with a Gaussian probability density function centered at zero and with identical widths,
\begin{equation}
g(x)=\frac{1}{\sqrt{2\pi\sigma^2}}\exp\left(\frac{-x^2}{2\sigma^2}\right), ~~~x=a,b,c,d,e,f.
\end{equation}
The probability that $x$ is between $x_l$ and $x_u$ ($x_l>x_u$) is defined to be $\int^{x_u}_{x_l}g(x){\rm d}x$.
The width $\sigma$ is defined such that the average absolute value of the trace of $M_1=m_1$. In the $2\times 2$ case, Eq.~(\ref{M1_2}),
$\sigma=\sqrt{\pi}/2$ (see below), while in the  $3\times 3$ case, Eq.~(\ref{M1_3}), $\sigma=\sqrt{\pi/6}$.

ii-Replace combinations of order one parameters by their {\sl average values} according to their probability density functions. For example
\begin{eqnarray}
\overline{|a+b|}&=&\int{|a+b|g(a)\cdots g(f)\rm d}a\cdots {\rm d}f=\int{|a+b|g(a)g(b)\rm d}a {\rm d}b, \\
&=&\frac{1}{2\pi\sigma^2}\int|a+b|\exp\left[\frac{-1}{4\sigma^2}\left((a+b)^2+(a-b)^2\right)\right]\left[\frac{{\rm d}(a+b){\rm d}(a-b)}{2}\right], \\
&=&\frac{\sqrt{\pi\sigma^2}}{2\pi\sigma^2}2\int_0^{\infty}\exp\left(-\frac{x^2}{4\sigma^2}\right)\frac{{\rm d}x^2}{2}, \\
&=&\frac{\sqrt{\pi\sigma^2}}{2\pi\sigma^2}4\sigma^2=\frac{2\sigma}{\sqrt{\pi}}.
\end{eqnarray}
In the $2\times 2$, $\overline{|a+b|}=1$ (by definition) while in the $3\times 3$ case,  $\overline{|a+b|}=\sqrt{2/3}$.

For other simple averages of sums or products, one can follow a similar procedure and obtain analytic results quickly. One other example is
\begin{eqnarray}
\overline{(a+b-2c)^2}&=&\frac{1}{(2\pi\sigma^2)^{3/2}}\int(a+b-2c)^2\exp\left[\frac{-1}{2\sigma^2}\left(\frac{1}{6}(a+b-2c)^2+\frac{1}{2}(a-b)^2+\frac{1}{3}(a+b+c)^2\right)\right]{\rm d}a {\rm d}b{\rm d}c,  \\
&=&\frac{2\sqrt{6}\pi\sigma^2}{(2\pi\sigma^2)^{3/2}}\int_{-\infty}^{\infty} x^2\exp\left(\frac{-x^2}{12\sigma^2}\right)\frac{{\rm d}x}{6}= \frac{\sqrt{6}\pi\sigma^2}{3(2\pi\sigma^2)^{3/2}}(6\sigma^2)^{3/2}\sqrt{2\pi},\\
&=&6\sigma^2.
\end{eqnarray}
Hence, in the $2\times 2$ case,  $\overline{(a+b-2c)^2}=3\pi/2$, as quoted in Sec.~\ref{2}.

Another, related, possibility would be not to compute average values, but instead identify ``confidence level ranges" for the various outputs as a function of the original width of the individual probability distributions, $\sigma$. For example, Eq.~(\ref{deltam_sol1}) could be reinterpreted as: the probability distribution for $\Delta m^2/m_0^2$ is centered at zero and has width $\sqrt2\sigma$. After defining the value of $\sigma$ (by requiring, say, that ${\rm Tr}(M_1)<m_1$ at the one-sigma confidence level), by requiring that the experimentally observed value is within some pre-defined confidence level, one can extract the value of $m_1/m_0$, as was done for the average-value procedure outlined above. Note that the requirement that $\overline{|a+b|}=1$ translates into a width $\sigma_{a+b}=\sqrt{2\pi}$. Hence, the probability that $a+b$ is between $-3$ and $3$ is around 77\%.


\begin{thebibliography}{99}
 
 \bibitem{SK_atm} Super-Kamiokande Coll. (Y.~Fukuda {\it et al.}), Phys. Rev. Lett. {\bf 81}, 1562 (1998).
 
 \bibitem{solar} B.T.~Cleveland {\it et al.}, Astroph. J. {\bf 496}, 505 (1998);
 GALLEX Coll. (W.~Hampell {\it et al.}), Phys. Lett. B {\bf 447}, 127 (1999);
 Super-Kamiokande Coll. (S.~Fukuda {\it et al}), Phys. Rev. Lett. {\bf 86}, 5656 (2001); 
 Sage Coll. (J.N.~Abdurashitov {\it et al.}), Zh. Eksp. Teor. Fiz. {\bf 122}, 211 (2002) [J. Exp. Theor. Phys. {\bf 95}, 181 (2002)];
 E.~Belotti for the GNO Coll., talk at the VIIIth International Conference on Topics in Astroparticle and Underground Physics (TAUP 03), Seattle, September 5--9, 2003.;
 SNO Coll. (Q.R~Ahmad {\it et al.}) Phys. Rev. Lett. {\bf 89}, 011301 (2002);
 SNO Coll. (S.N.~Ahmed {\it et al.}) nucl-ex/0309004, and many references therein.

 \bibitem{KamLAND} KamLAND Coll. (K. Eguchi {\it et al.}), Phys. Rev. Lett. {\bf 90} 021802 (2003) 

\bibitem{PDG} Particle Data Group (K.~Hagiwara {\it et al.}) Phys. Rev. D {\bf 66}, 010001(2002).

\bibitem{anarchy_stat}  A.~de Gouv\^ea and H.~Murayama, Phys.\ Lett.\ B {\bf 573}, 94 (2003).


\bibitem{CHOOZ} CHOOZ Coll. (M.~Apollonio {\it et al.}) Phys. Lett. {\bf B} 466, 415 (1999); Eur. Phys. J. C {\bf 27}, 331 (2003).

\bibitem{CHOOZ_bound} G.L.~Fogli {\it et al.}, hep-ph/0308055 for the most recent bound, including the revised SuperKamiokande estimate for the atmospheric mass-squared difference \cite{SuperK_new}.

\bibitem{SU(5)_anarchy} see, for example, T.~Moroi, JHEP {\bf 0003}, 019 (2000).
See also D.~Chang, A.~Masiero and H.~Murayama, Phys.\ Rev.\ D {\bf 67}, 075013 (2003). 

\bibitem{minimal_model} See, for example, A.~de Gouv\^ea and J.W.F.~Valle, Phys. Lett. B {\bf 501}, 115 (2001).

\bibitem{anarchy} L.J.~Hall, H.~Murayama and N.~Weiner, Phys.\ Rev.\ Lett.\  {\bf 84}, 2572 (2000).

\bibitem{superbeam} Y.~Itow {\it et al.}, hep-ex/0106019;
D.~Ayres {\it et al.}, hep-ex/0210005;
M.V.~Diwan {\it et al.}, Phys.\ Rev.\ D {\bf 68}, 012002 (2003).
For a recent review of current proposals, see T.~Nakaya, Nucl.\ Phys.\ Proc.\ Suppl.\  {\bf 118}, 210 (2003).

\bibitem{new_reactor} H.~Minakata {\it et al.}, Phys.\ Rev.\ D {\bf 68}, 033017 (2003); M.H.~Shaevitz and J.M.~Link, hep-ex/0306031; K.~Heeger, talk at the Weak Interactions and Neutrinos Workshop -2003 (WIN 03), Lake Geneva, October 6--11, 2003. See also 
P.~Huber, M.~Lindner, T.~Schwetz and W.~Winter, Nucl.\ Phys.\ B {\bf 665}, 487 (2003).


\bibitem{models} For recent reviews see R.N.~Mahapatra, hep-ph/9910365;
S.M.~Barr and I.~Dorsner, Nucl.\ Phys.\ B {\bf 585}, 79 (2000);
G.~Altarelli and F.~Feruglio, hep-ph/0206077; hep-ph/0306265;
S.F.~King, hep-ph/0310204; and references therein. 

\bibitem{GUT_models} See, for example, H.S.~Goh, R.N.~Mohapatra and S.P.~Ng, hep-ph/0308197; 
C.H.~Albright and S.M.~Barr, Phys.\ Rev.\ D {\bf 67}, 013002 (2003);
H.~D.~Kim and S.~Raby, JHEP {\bf 0307}, 014 (2003);
M.~C.~Chen and K.~T.~Mahanthappa, Phys.\ Rev.\ D {\bf 68}, 017301 (2003) for some recent efforts.

\bibitem{osc_anal} see, for example, M.~Maltoni, T.~Schwetz, M.A.~Tortola and J.W.F.~Valle, hep-ph/0309130 for the most  recent attempt at fitting all solar, reactor, and atmospheric data, and many references therein.

\bibitem{aguilar_saavedra} J.A.~Aguilar-Saavedra, Phys.\ Rev.\ D {\bf 67}, 073026 (2003);
J.A.~Aguilar-Saavedra, G.C.~Branco and F.R.~Joaquim, hep-ph/0310305.


\bibitem{srhnd} S.F.~King, Phys.\ Lett.\ B {\bf 439}, 350 (1998);
Nucl. Phys. B {\bf 562}, 57 (1999);
S.~Davidson and S.F.~King, Phys.\ Lett.\ B {\bf 445}, 191 (1998).

\bibitem{SuperK_new} Y.~Hayato for the Super-Kamiokande Coll., talk at the International European Conference on High Energy Physics (HEP 2003), Aachen, Germany, 2003. [http://eps2003.physik.rwth-aachen.de]. 

\bibitem{dark_side} A.~de Gouv\^ea, A.~Friedland and H.~Murayama, Phys.\ Lett.\ B {\bf 490}, 125 (2000).

\bibitem{tri-bimaximal} P.F.~Harrison, D.H.~Perkins, and W.G.~Scott, Phys.\ Lett.\ B {\bf 530}, 167 (2002);
P.F.~Harrison and W.G.~Scott, Phys.\ Lett.\ B {\bf 535}, 163 (2002).



\bibitem{bi-maximal} F.~Vissani, hep-ph/9708483;
V.~Barger, S.~Pakvasa, T.~Weiler, and K.~Whisnant, Phys.\ Lett.\ B {\bf 437}, 107 (1998);
A.~Baltz, A.S.~Goldhaber, and M.~Goldhaber, Phys.\ Rev.\ Lett. {\bf 81} 5730 (1998);
G.~Altarelli and F.~Feruglio, Phys.\ Lett.\ B {\bf 439}, 112 (1998);
M.~Jezabek and Y.~Sumino, Phys.\ Lett.\ B {\bf 440}, 327 (1998).

\bibitem{a4} E.~Ma and G.~Rajasekaran, Phys.\ Rev.\ D {\bf 64}, 113012 (2001);
E.~Ma, Mod.\ Phys.\ Lett.\ A {\bf 17}, 2361 (2002);
K.~Babu, E.~Ma, and J.W.F.~Valle, Phys.\ Lett.\ B {\bf 552} 207 (2003).

\bibitem{MINOSetal} G.~Tzanakos, talk at the 5th International Workshop on Neutrino Factories \& Superbeams, New York,  5--11 June 2003, http://www.cap.bnl.gov/nufact03/agenda.xhtml.;
W.C.~Louis, talk at the 5th International Workshop on Neutrino Factories \& Superbeams, New York,  5--11 June 2003, http://www.cap.bnl.gov/nufact03/agenda.xhtml              

\bibitem{bernstein} R.~Bernstein, talk at the 5th International Workshop on Neutrino Factories \& Superbeams, New York,  5--11 June 2003, http://www.cap.bnl.gov/nufact03/agenda.xhtml. 

\bibitem{atmospheric_new} T.~Tabarelli de Fatis, Nucl.\ Phys.\ Proc.\ Suppl.\  {\bf 118}, 118 (2003).

\bibitem{peres_smirnov} O.L.G.~Peres and A.Yu.~Smirnov, hep-ph/0309312.

\bibitem{nufact} see, for example, C.~Albright {\it et al.}, hep-ex/0008064;
M.~Apollonio {\it et al.}, hep-ph/0210192.

\bibitem{Grimus_Lavoura} W.~Grimus and L.~Lavoura, JHEP {\bf 0107}, 045 (2001).

\bibitem{afm} G.~Altarelli, F.~Feruglio and I.~Masina, JHEP {\bf 0301}, 035 (2003).

\bibitem{Petcov_new} P.H.~Frampton, S.T.~Petcov, and W.~Rodejohann, hep-ph/0401206.

 \end{thebibliography}
 \end{document}